\documentclass{amsart}      
\usepackage{amssymb}      
\usepackage{latexsym}      
      
\newcommand{\ZZ}{{\mathbb Z}}      
\newcommand{\RR}{{\mathbb R}}      
\newcommand{\CC}{{\mathbb C}}      
\newcommand{\NN}{{\mathbb N}}      
      
\newcommand{\aA}{{\mathbb A}}   
\newcommand{\QQ}{{\mathbb Q}}  
\newcommand{\sS}{{\mathbb S}}

\newtheorem{theorem}{Theorem}[section]         
\newtheorem{remark}[theorem]{Remark}           
\newtheorem{lemma}[theorem]{Lemma}    
\newtheorem{prop}[theorem]{Proposition}         
\newtheorem{coro}[theorem]{Corollary}         
\newtheorem{definition}[theorem]{Definition}         
\newcommand{\tr}{{\mathrm{tr}}}      
  
\newcommand{\CalG}{\mathcal{G}}  
\newcommand{\CalGomega}{\mathcal{G}(\Omega,T)}  
  
\newcommand{\CalN}{\mathcal{N}(\Omega,T, \mu)}  
\newcommand{\CalA}{\mathcal{A} (\Omega,T)}  
\newcommand{\CalP}{\mathcal{P}}  
\newcommand{\CalX}{\mathcal{X}}  
  
\newcommand{\CalF}{\mathcal{F}}  
\newcommand{\CalV}{\mathcal{V}}  
  
\newcommand{\CalM}{\mathcal{M}}  
\newcommand{\CalB}{\mathcal{B}}  
\newcommand{\CalK}{\mathcal{K}}  
\newcommand{\CalT}{\mathcal{T}}

\newcommand{\Oomega}{(\Omega,T)}

\begin{document}    
\title[Random operators associated to Delone sets]{Delone dynamical systems   
and   
associated random operators}

\author[D. Lenz, P. Stollmann]{Daniel Lenz~$^{2}$,\  
Peter Stollmann~$^{3}$}

\begin{abstract} We carry out a careful study of  basic   
topological and ergodic   
features of Delone dynamical systems. We then investigate the associated   
topological groupoids and in particular their representations on certain   
direct integrals with non constant fibres. Via non-commutative-integration   
theory these representations give rise to von Neumann algebras of random   
operators. Features of these algebras and operators  are discussed.   
Restricting our attention to a certain subalgebra of tight binding   
operators, we then discuss a  Shubin trace formula.   
  
\end{abstract}  
  
\maketitle  \vspace{0.3cm}   
\noindent $^{2,}$\footnote{Research partly supported  
by the DFG in the priority program Quasicrystals} Fakult\"{a}t f\"{u}r   
Mathematik, Technische   
Universit\"{a}t Chemnitz,   
D-09107 Chemnitz, Germany;  
E-mail:  D.Lenz@mathematik.tu-chemnitz.de,\\[0.1cm]  
  
\noindent  
$^{3,1}$  Fakult\"{a}t f\"{u}r Mathematik, Technische Universit\"{a}t   
Chemnitz,   
D-09107 Chemnitz, Germany; E-mail:  P.Stollmann@mathematik.tu-chemnitz.de

\section*{Introduction}  
  
The study of disorder is one of the most important issues today. In  
mathematical models of solid state physics order or disorder are  
expressed in terms of the hamiltonian that drives the system. The  
latter operator appears in the Schr\"odinger equation and its  
properties are used to describe the electronic properties of the solid  
under consideration.  
  
Both order and disorder can be cast in the framework of parametrized  
operators.  Let us be a little more precise on that point. We start at  
one extreme: a perfectly orderered solid, an ideal crystal. Since the  
atomic positions belong to a lattice $\Gamma$ in this case, the  
hamiltonian exhibits a corresponding translational symmetry. This  
symmetry is encoded in the quotient $X/\Gamma$ of the underlying space  
$X\in\{\RR^d,\ZZ^d\}$ which leads to a family of operators  
parametrized by $X/\Gamma$. As elementary as this observation is, it  
leads to a number of important consequences. One of these consequences  
concerns the nature of the spectrum and the dynamics of the  
hamiltonian: periodic operators (at least under some additional  
assumptions) have purely absolutely continuous spectrum and only  
extended states.  The other extreme case concerns models that are  
statistically independent at distant parts of space. Important and  
well studied are models of Anderson type for which the parameter space  
is typically of the form $I^{\ZZ^d}$, $I$ a compact interval in $\RR$.  
For these models a completely different spectral and dynamical picture  
is expected with energy regions filled by dense pure point spectrum  
with localized eigenfunctions and absence of mobility. The key notion  
is localization, a phenomenon that is rigorously proved in certain  
cases, see \cite{cl,pf,s}. The models we want to study are situated in  
between ordered (e.g. periodic) and heavily disordered (e.g. Anderson  
models) and are used to describe quasicrystals. The discovery of  
quasicrystals (see the celebrated 1984 paper \cite{she}) has led to an  
impressive research activity. However, mathematically rigorous results  
concerning the spectral theory and dynamics of quasicrystalline  
Schr\"{o}dinger operators are somewhat rare except for one dimensional  
papers.   
  
In the present paper we report on some basic  
issues concerning the treatment of the  
multidimensional case. Namely, we describe how Delone dynamical systems   
of finite type can be used as the relevant parameter  
spaces.  It turns out that  
there are aspects common with the one dimensional situation: local  
finiteness and a hierarchical order.  There is one important  
difference as well: a more complicated geometry and the absence of a  
lattice structure. This leads to complications concerning the  
parametrized family of hamiltonians that appears. The latter are  
defined on different Hilbert spaces. The first point we mentioned is  
reflected in the strong ergodic properties that are the same as in the  
one dimensional case. The second point leads to the lack of spectral  
consequences that follow from these ergodic properties.  
  
At this point let us end this general preview and refer the reader to  
the main text for more details. Instead we go on to relate our results  
to what can be found in the literature:  
  
Various features of Delone sets and tilings have been considered in  
the literature. See, e.g., \cite{BHZ,kp,l1,l2,lp,lms,se,sol1,sol2} and the  
literature cited there.  However, as will be seen in Section 1 below  
there are certain misunderstandings concerning the problem of  
defining a suitable topology on the set of Delone  
sets.  
  
On the other hand, a thorough study of ``almost random operators'' in  
an algebraic setting focusing on $K$-theory can be found in  
\cite{be1,be2,be3}. This has been taken up by Kellendonk who  
introduced certain $C^\ast$-algebras associated to tilings and studied  
their $K$-theory \cite{ke,ke2}, see \cite{Ap,kp,put} as well.  In  
fact, we have been inspired by these works although they do not cover  
all the results mentioned here. Our focus is rather on general  
features of random operators associated to tilings. These features  
include (almost sure) constancy of the spectrum and its various parts,  
absence of discrete spectrum and validity of a Shubin trace formula.  
  
The important method here is Connes noncommutative integration theory,  
which we use to associate a von Neumann algebra to Delone dynamical  
systems. We should stress that this does not mean to associate in some  
general fashion a von Neumann algebra to a dynamical system. Rather we  
use the specific situation at hand as well as the guidance provided by  
the physical background.\\[.3cm]  
  
\noindent{\bf Acknowledgement.}  
It is our pleasure to acknowledge fruitful correspondence with J. Lagarias  
and B. Solomyak.

\section{Generalities on Delone sets}  
  
The aim of this section is to recall standard concepts from the theory  
of Delone sets and to introduce a suitable topology on the closed sets  
in euclidian space.  
  
\medskip

We will be concerned with Delone sets in $\RR^d$. A subset $\omega$ of  
$\RR^d$ is called a {\it Delone set} if there exist $0<r,R<\infty$  
such that $r \leq \|x-y\|$ whenever $x,y\in \omega$ with $x\neq y$,  
and $B_R (x)\cap \omega \neq \emptyset$ for all $x\in \RR^d$.  Here,  
the Euclidean norm on $\RR^d$ is denoted by $\|\cdot\|$ and $B_s (x)$  
denotes the (closed) ball in $\RR^d$ around $x$ with radius $s$. The  
set $\omega$ is then also called an $(r,R)$-set.  We will be  
particularly interested in the restrictions of Delone sets to bounded  
sets. In order to treat these restrictions, we introduce the following  
definition.

\begin{definition}{\rm (a)} A pair  $(\Lambda,Q)$ consisting of a bounded    
  subset $Q$ of $\RR^d$ and $\Lambda\subset Q$ finite is called a {\rm  
    pattern}.  
  The set $Q$ is called the {\rm support of the pattern}. \\  
  {\rm (b)} A pattern $(\Lambda,Q)$ is called a {\rm ball pattern} if  
  $Q=B_s (x)$ with $x\in \Lambda$ for suitable $x\in \RR^d$ and $s\in  
  (0,\infty)$.  
\end{definition}

The pattern $(\Lambda_1,Q_1)$ is contained in the pattern  
$(\Lambda_2,Q_2)$ written as $(\Lambda_1,Q_1) \subset (\Lambda_2,Q_2)$  
if $Q_1\subset Q_2$ and $\Lambda_1=Q_1\cap \Lambda_2$.  Diameter,  
volume etc. of a pattern are defined to be the diameter, volume etc of  
its support.  For patterns $X_1=(\Lambda_1,Q_1) $ and $X_2=  
(\Lambda_2,Q_2)$, we define $\sharp_{X_1} X_2$, {\it the number of  
  occurences of $X_1$ in $X_2$}, to be the number of elements in  
$\{t\in \RR^d : \Lambda_1 +t \subset \Lambda_2, Q_1 +t \subset Q_2\}$.  
  
For further investigation we will have to identify patterns which are  
equal up to translation. Thus, on the set of patterns we introduce an  
equivalence relation by setting $(\Lambda_1,Q_1)\sim (\Lambda_2, Q_2)$  
if and only if there exists a $t\in \RR^d$ with $\Lambda_1 = \Lambda_2  
+ t$ and $Q_1=Q_2 + t$.  In this latter case we write  
$(\Lambda_1,Q_1)= (\Lambda_2, Q_2)+ t$.  The class of a pattern  
$(\Lambda,Q)$ is denoted by $[(\Lambda,Q)]$.  The notions of diameter,  
volume, occurence etc. can easily be carried over from patterns to  
pattern classes.

Every Delone set $\omega$ gives rise to a set of pattern classes,  
$\CalP (\omega)$ viz $\CalP (\omega)=\{ Q\wedge \omega :  
Q\subset\RR^d\: \mbox{bounded and measurable} \}$, and to a set of  
ball pattern classes $\CalP_B (\omega)) =\{ B_s (x)\wedge \omega :  
x\in \omega, s>0\}$.  Here we set $Q\wedge \omega= [(\omega \cap Q,  
Q)]$.  
  
For $s\in (0,\infty)$, we denote by $\CalP_B^s (\omega)$ the set of  
ball patterns with radius $s$; note the relation with $s$-patches as 
considered in \cite{l1}.  A Delone set is said to be of {\it  
  finite type} if for every radius $s$ the set $\CalP_B^s (\omega)$ is  
finite. We refer the reader to \cite{l1} for a detailed discussion of   
Delone sets of finite type.  
  
Next we introduce a suitable topology on the set of closed subsets of  
$\RR^d$.  Although it is basically known how this can be done, we will  
take some care.  Actually, it turns out that certain statements in  
\cite{l2,sol2} concerning this issue are, while morally true, not  
completely correct ;-)  
  
Denote by $\CalF (\RR^d)$ the set of closed subsets of $\RR^d$ and  
recall that there is a natural action $T$ of $\RR^d$ on $\CalF  
(\RR^d)$ given by $T_t G= G+t$. We aim at a topology on $\CalF  
(\RR^d)$ that fulfills two requirements: the action $T$ should be  
continuous and two sets that are close to each other with respect to  
the topology are supposed to be such that their finite parts have  
small {\it Hausdorff distance}. The latter can be defined by  
$$  
d_H(K_1,K_2):= \inf( \{ \epsilon >0: K_1\subset  
U_\epsilon(K_2)\wedge K_2\subset U_\epsilon(K_1)\}\cup \{1\} ),  
$$  
where $K_1,K_2$ are compact subsets of a metric space $(X,d)$  
and $U_\epsilon(K)$ denotes the open $\epsilon$-neighborhood around  
$K$. The extra $1$ is to deal with the empty set that is included in  
$\CalK (X):=\{ K\subset X: K \mbox{ compact}\}$. It is wellknown that  
$(\CalK (X), d_H)$ is complete if $(X,d)$ is complete and compact  
if $(X,d)$ is compact.  Quite often, the Hausdorff distance is  
defined between nonvoid compact sets only. The way we defined it,  
$\emptyset$ is added as an isolated point. For our purposes later on  
it will be important to have the empty set at our disposal and it will  
no longer be an isolated point.  
  
A natural first attempt to define a suitable topology goes as follows.  
Abbreviate $B_r (0)=:B_r$ and $\CalK (B_r):= \CalK_r$ which is a  
compact metric space by what we just mentioned. We call the initial  
topology on $\CalF (\RR^d)$, induced by the restriction mappings  
$$  
J_R:\CalF (\RR^d)\to \CalK_R, F\mapsto F\cap B_R, R>0  
$$  
the {\it topology of local Hausdorff convergence}. Of course this  
topology satisfies the second requirement listed above.   
However a serious problem connected with that topology comes from the   
fact, that  
$F$ and $G$ might be closer in Hausdorff distance than $F\cap B_k$ and  
$G\cap B_k$ as two relatively close portions of $F$ and $G$ might just  
lie inside respectively outside the ball $B_R$. Put differently, the  
restriction $F\mapsto F\cap B_r$ is by no way a contraction from  
$\CalK_R$ to $\CalK_r$.  In particular, $T$ does not act continuously.  
Consider, e.g., $F:=\ZZ\in \CalF (\RR)$; for any $t\in (0,1)$ we get  
that  
$$  
d_H((T_tF)\cap B_k,F\cap B_k)=\max\{ |t|, 1-|t|\} .  
$$  
Consequently, for any $k\ge 3$ and $t\in (0,1)$, $T_tF\not\in  
V_k(F)$.  To circumvent this lamentable fact we introduce  
  
$$  
d_k(F,G):= \inf( \{ \epsilon >0: F\cap B_k\subset  
U_\epsilon(G)\wedge G\cap B_k\subset U_\epsilon(F)\}\cup \{1\} ),  
$$  
a measure for the distance of $F$ and $G$ that is monotone in the  
cutoff parameter $k$, i.e., $d_k(F,G)\le d_{k+1}(F,G)$. Moreover,  
$d_k(F,G)\le d_H(F\cap B_k,G\cap B_k)$. Unfortunately, the $d_k$ do  
not satisfy the triangle inequality. To see this, consider, in $\RR$,  
the case $k=1$ and the sets $F=\{1-\epsilon\} , G=\{1+\epsilon\},  
H=\emptyset $, leading to $d_k(F,G)= 2 \epsilon$, $d_k(G,H)=0$ but  
$d_k(F,H)=1$ (if $\epsilon$ is small enough).

We use them to define a topology coarser than the topology of local  
Hausdorff convergence via the neighborhood basis  
$$  
U_{\epsilon,k}(F):= \{ G\in\CalF (\RR^d): d_k(F,G)\le\epsilon \},  
\epsilon >0,k\in \NN ;  
$$  
we call the corresponding topology $\tau_{nat}$ the {\it natural topology}. By  
the definition of $d_k$ it is clear that $d_k(T_tF,F)\le |t|$ for any  
$F\in \CalF (\RR^d), t\in \RR^d$ so that translations are continuous.  
Fortunately this topology has nice compactness properties as seen in:  
  
\begin{theorem} $\CalF (\RR^d)$ endowed with the natural topology    
$\tau_{nat}$ is compact.  
\end{theorem}  
  
\begin{proof} Let $(F_\iota)_{\iota\in I}$ be a net in $\CalF (\RR^d)$.  
We use that $\CalK_{k+\frac12}$ is a compact metric space for every  
$k$. Therefore, $(F_\iota\cap B_{k+\frac12})_ {\iota\in I}$ has an  
accumulation point $C_{k+\frac12}$ in $\CalK_{k+\frac12}$ for every  
$k$ and we find a subsequence converging in the Hausdorff metric. By a  
standard diagonal argument we can arrange a common subsequence, call  
it $(F_m)_{m\in\NN}$, such that $(F_m\cap B_{k+\frac12})_{m\in\NN}$  
converges to $C_{k+\frac12}$ for every $k\in\NN$;  
$$C := \bigcup_{k\in \NN}C_{k+\frac12} $$  
is a good candidate for the  
limit of the $(F_m)$ in the natural topology. Let us first note that  
$$  
C\cap B_k= C_{k+\frac12}\cap B_k\qquad (*) .  
$$  
(This is clear but it is here that the $\ldots+\frac12$ in the  
definitions above gets important. It is not true in general that  
$C\cap B_{k+\frac12}= C_{k+\frac12}\cap B_{k+\frac12}$. We will meet  
this kind of effect again later when discussing the lack of  
compactness for the topology of local Hausdorff convergence in the  
paragraph following the present proof.) From $(*)$ it follows that $C$  
is closed. We have to show that  
$$  
d_k(F_m,C)\to 0\mbox{ as }m\to\infty  
$$  
for any $k$; so fix $k$. If  
$$  
1 >\delta_m > d_H(F_m\cap B_{k+\frac12},C_{k+\frac12})  
$$  
we get that  
$$  
C\cap B_k\subset C_{k+\frac12}\subset U_{\delta_m}(F_m\cap  
B_{k+\frac12})  
$$  
by definition of the Hausdorff metric. Conversely,  
$$  
F_m\cap B_k\subset F_m\cap B_{k+\frac12} \subset  
U_{\delta_m}(C_{k+\frac12}) \subset U_{\delta_m}(C) .  
$$  
Put together, we get  
$$  
d_k(F_m,C)\le \delta_m  
$$  
for $\delta_m > d_H(F_m\cap B_{k+\frac12},C_{k+\frac12}) $, i.e.,  
$$  
d_k(F_m,C)\le d_H(F_m\cap B_{k+\frac12},C_{k+\frac12})  
$$  
and the latter tends to $0$ as $m\to\infty$. Thus we have proved  
that every net in $\CalF (\RR^d)$ has a converging subsequence.  
\end{proof}  
   
\medskip  
  
Note that, interestingly, no additional  
properties are needed for compactness. Of course, this result immediately   
gives compactness of certain subsets of the set of all Delone sets, e.g.   
compactness of the union over $R$ of the $(r,R)$-sets for any fixed  
value of $r$.  
  
One might think that the  
topology of local Hausdorff leads to a compact space as well, since in  
this topology, $\CalF (\RR^d)$ is considered as a subset of the  
product of the $\CalK_R$, which is compact by Tychonovs theorem.  
However, it is not closed, as seen for $F_n:=\{ 1+\frac1n \}$; of  
course, $R=1$ is the crucial value. This latter sequence also shows  
directly that the topology of local Hausdorff convergence is not  
compact and not too natural either ;-)  
  
We will also use the natural topology to define a topology on tiling  
spaces in a quite general setting.

Let us note in passing that the metric $\rho$ proposed in \cite{sol2}  
as well as the metric in \cite{l2}  
do not satisfy the triangle inequality and that these metrics are   
restricted to Delone sets. This is again due to the  
phenomenon alluded to above, namely that restricting sets does not  
make the Hausdorff distance smaller.  See, however,  \cite{lms}, in which a  
metric on the set of Delone sets is constructed. A discussion in a more 
general framework can be found in \cite{schlo}, where the author constructs a 
topology on the set of closed discrete subsets of a locally compact 
$\sigma$-compact space. In the case of $\RR^d$ this topology coincides with 
the restriction of the above given
natural topology.

To settle the issue of metrizability of the natural topology we next mention  
an alternative approach.    
The method we are going to outline now has most definitely  been pointed  
out to us by someone else. Unfortunately, we were not able to find out 
by whom.  
  
We use the stereographic projection to identify points $x\in\RR^d  
\cup\{\infty\}$ in the one-point-compactification of $\RR^d$ with  
the corresponding points $\tilde{x}\in\sS^d$. Clearly, the latter  
denotes the $d$-dimensional unit sphere $\sS^d=\{\xi\in\RR^{d+1}:  
\|\xi\| =1\}$. Now $\sS^d$ carries the euclidean metric $\rho$. Since  
the unit sphere is compact and complete, we can associate a complete metric  
$\rho_H$ on $\CalK(\sS^d)$ by what we said above.  
  
For $F\in \CalF(\RR^d)$ write $\tilde{F}$ for the corresponding  
subset of $\sS^d$ and define  
$$  
\rho(F,G):=   
\rho_H(\widetilde{F\cup\{\infty\}},\widetilde{G\cup\{\infty\}})\mbox{  for  }  
F,G\in \CalF(\RR^d) .  
$$  
Although this constitutes a slight abuse of notation it makes sense since  
$\widetilde{F\cup\{\infty\}},\widetilde{G\cup\{\infty\}}$ are compact in   
$\sS^d$ provided  
$F,G$ are closed in $\RR^d$.  
  
We have the following result:  
  
\begin{prop} The metric $\rho$ above induces the natural topology  
on $\CalF(\RR^d)$.  
\end{prop}  
\begin{proof}  
An explicit calculation of the stereographic projection shows that  
\begin{equation}\label{eq1}  
\rho(\widetilde{x},\widetilde{\infty})\le \frac{42}{\| x\|}  
\end{equation}  
for $x\in\RR^d$ as well as  
\begin{equation}\label{eq2}  
\rho(\widetilde{x},\widetilde{y})\le 2\| x-y\| .  
\end{equation}  
We want to show that the identity $id:(\CalF(\RR^d),\tau_{nat})\to  
(\CalF(\RR^d),\rho)$ is continuous. Since $(\CalF(\RR^d),\tau_{nat})$  
is a compact space this implies that $id:(\CalF(\RR^d),\tau_{nat})\to  
(\CalF(\RR^d),\rho)$ is in fact a homeomorphism. To prove the desired  
continuity, fix $F\in \CalF(\RR^d)$ and $\varepsilon >0$. We have to find  
a basic neighborhood $U_{\delta,k}$ that is contained in the $\varepsilon$-  
ball around $F$ (with respect to $\rho$, of course). To this end  
choose $k\in\NN$ such that $42/k < \varepsilon$ and $\delta=\varepsilon/2$.  
  
This is a good choice; in fact, let $G\in U_{\delta,k}$.   
Combining $F\cap B_k\subset U_\delta(G)$ with (\ref{eq2}) we get that  
\begin{equation}\label{eq3}  
\widetilde{F\cap B_k}\subset U^\rho_\varepsilon(\widetilde{G}) ,  
\end{equation}  
where the superscript $\rho$ indicates the underlying metric space.  
By (\ref{eq1}) and the choice of $k$ we know that  
\begin{equation}\label{eq4}  
\widetilde{F\cap (B^c_k\cup\{\infty\}})\subset   
U^\rho_\varepsilon(\widetilde{G\cup\{\infty\}}) .  
\end{equation}  
Relations (\ref{eq3}) and (\ref{eq4}) together yield that  
$$  
\widetilde{F\cup\{\infty\}}\subset U^\rho_\varepsilon  
(\widetilde{G\cup\{\infty\}}) .  
$$  
The corresponding relation with $F$ and $G$ interchanged follows in the  
same way so that  
$$  
\rho_H(\widetilde{F\cup\{\infty\}},(\widetilde{G\cup\{\infty\}})\le   
\varepsilon .  
$$  
This proves the asserted continuity.  
\end{proof}  
Next, we define Delone dynamical systems, following \cite{lp}   
and single out some important  
properties:  

\begin{definition}  {\rm (a)} Let $\Omega$ be a set of Delone sets.  
  The pair $(\Omega,T)$ is called a {\em Delone dynamical system}  
  (DDS) if $\Omega$ is invariant under the shift $T$ and closed in the  
  natural  
  topology. \\  
  {\rm (b)} A DDS $\Oomega$ is said to be of {\em finite type} if  
  $\cup_{\omega\in \Omega} P_B^s (\omega)$ is finite for every $s>0$. \\  
  {\rm (c)} Let $0<r,R<\infty$ be given.  A DDS $\Oomega$ is said to  
  be an  
  $(r,R)$-{\em system} if every $\omega\in \Omega$ is an $(r,R)$-set.\\  
  {\rm (d)} The set $\CalP (\Omega)$ of {\em pattern classes  
    associated to a DDS} $\Omega$ is defined by $\CalP  
  (\Omega)=\cup_{\omega\in \Omega} \CalP (\omega)$.  
\end{definition}

\begin{remark}{\rm (a) Whenever $\Oomega$ is a Delone dynamical system,   
    there exists an $R>0$ with $B_R (x)\cap \omega\neq \emptyset$ for  
    every $\omega\in \Omega$ and every $x\in \RR^d$. This follows  
    easily as $\Omega$  
    is closed and invariant under the action of $T$.\\  
    (b) Every DDSF is an $(r,R)$-system for suitable $0<r,R<\infty$.\\  
    (c) Let $\omega$ be an $(r,R)$-set and let $\Omega_\omega$ be the  
    closure of $\{T_t \omega : t\in \RR^d\}$ in $\CalF (\RR^d)$ with  
    respect to the natural topology.  Then $(\Omega_\omega,T)$ is an  
    $(r,R)$-system.  }  
\end{remark}

For a DDSF, there is a simple way to describe convergence in the natural  
topology. This is shown in the following lemma. We omit the  
straightforward proof.  
  
\begin{lemma} If $\Oomega$ is a DDSF then a sequence $(\omega_n)$   
  converges to $\omega$ in the natural topology if and only if there  
  exists a sequence $(t_n)$ converging to $0$ such that for every  
  $L>0$ there is an $n_0\in\NN$ with $(\omega_n+t_n)\cap  
  B_L=\omega\cap B_L$ for $n\geq n_0$.  
\end{lemma}

We will need standard notions from the theory of dynamical systems.  
Namely, a DDS is called {\it minimal} if every orbit is dense.   
It is  
called {\it uniquely ergodic} if there exists exactly one  
$T$-invariant probability measure. It is called {\it aperiodic} if  
$T_t$ does not have a fixed point for any $t\neq 0$.  
  
For $s>0$ and $Q\in \RR^d$, we denote by $\partial_s Q$ the set of  
points in $\RR^d$ whose distance to the boundary of $Q$ is less than  
$s$. A sequence $(Q_n)$ of bounded subsets of $\RR^d$ is called a van  
Hove sequence if $|Q_n|^{-1} |\partial_s Q_n|\longrightarrow 0,  
n\longrightarrow 0$ for every $s>0$.

\begin{theorem} \label{einderg} Let $\Oomega$ be a DDSF. Then $\Oomega$ is uniquely ergodic   
  if and only if, for every pattern class $P$ the frequency  
  $\lim_{n\to \infty} |Q_n|^{-1} \sharp_{P} (\omega\wedge Q_n)$ exists  
  uniformly in $\omega \in \Omega$ for every van Hove sequence $Q_n$.  
\end{theorem}  
  
\begin{remark}{\rm After a first version of the present paper was on the web,  
B. Solomyak kindly informed us of the work \cite{lms} by Lee, Moody and  
Solomyak. As \cite{lms}, Theorem 2.7 the reader can find a result analogous to the preceding Theorem.}  
\end{remark}  
   
\begin{proof} For the ``if'' part we can refer to \cite{sol1}, Theorem  
3.3.  There, it is additionally assumed that the tiling dynamical  
system has the local isomorphism property. The latter is only used,  
however, to guarantee that frequencies are strictly positive. We will  
describe how to pass from Delone dynamical systems to tilings in  
detail later.  
  
For the ``only if'' part we construct a continuous function on  
$\Omega$ that essentially counts occurences of a fixed pattern. Here  
are the details:  
  
Fix $0<r<R$ such that $\Omega$ is an $(r,R)$-system. We fix a pattern  
$P=(\Lambda_P,Q_P)$ with $0\in\Lambda_P$ and diameter $R_P$. Moreover,  
we choose an auxiliary function $g\in C_c(\RR^d)$, $g\ge 0$ with  
support contained in $B_{\frac{r}{42}}$ and normalized to $\int g(x)dx  
=1$. With the help of $g$ we define  
$$  
f_P:\Omega\to \RR, f_P(\omega)=g(t)\mbox{ iff }Q_P\wedge  
(\omega+t)=P  
$$  
(where such a $t$ is uniquely determined in case $g(t)\not = 0$)  
and $0$ if no translate of $P$ appears in $\omega$. We consider all  
the points of $\omega$ at which a copy of $P$ is centered, namely  
$$  
S_\omega:=\{ s_n\in\RR^d :Q_P\wedge (\omega+s_n)=P\}  
$$  
and remark that the distinct points of $S_\omega$ have distance at  
least $r$ since $\Omega$ is an $(r,R)$-system. Using the set  
$S_\omega$, we get  
$$  
f_P(\omega +t)=\sum_ng(s_n-t)  
$$  
and, therefore,  
\begin{eqnarray*}  
\int_Qf_P(\omega+t)dt &=& \sum_n\int_Q g(s_n-t)dt\\  
&=&\sum_{B_{\frac{r}{42} (s_n)}\cap Q\not=\emptyset}  
                                 \int_Qg(s_n-t)dt  
\end{eqnarray*}  
Thus, if we let  
$$  
Q^-:=\{ y\in Q:y+B_{R_P+r}\subset Q\}, Q^+:= U_r(Q),  
$$  
we get that  
$$  
\sharp_P Q\wedge\omega= \#\{n: s_n+Q_P\subset Q\}  
$$  
along with the inequalities  
\begin{eqnarray*}  
\int_{Q^-}f_P(\omega+t)dt &\le& \#\{n: B_{\frac{r}{42}}(s_n)\cap   
Q^-\not=\emptyset\}\\  
&\le& \sharp_P Q\wedge\omega= \#\{n: s_n+Q_P\subset Q\}\\  
&\le& \#\{n: B_{\frac{r}{42}}(s_n)\subset Q^+\}\\  
&\le& \int_{Q^+}f_P(\omega+t)dt .  
\end{eqnarray*}  
Now, take a van Hove sequence $(Q_n)$. Then, since $\Oomega$ is  
uniquely ergodic,  
\begin{equation*}  
\lim_{n\to\infty}\frac{1}{|Q_n|}\int_{Q^-_n}f_P(\omega+t)dt=  
\lim_{n\to\infty}\frac{1}{|Q_n|}\int_{Q_n}f_P(\omega+t)dt  
= \int_\Omega f_P(\omega)d\mu(\omega) ,  
\end{equation*}  
where $\mu$ is the unique normalized invariant measure. Similarly,  
$$  
\lim_{n\to\infty}\frac{1}{|Q_n|}\int_{Q^+_n}f_P(\omega+t)dt=  
\int_\Omega f_P(\omega)d\mu(\omega).$$  
Using the inequality above, we  
now find that  
$$  
\frac{1}{|Q_n|}\int_{Q^-_n}f_P(\omega+t)dt \le  
\frac{1}{|Q_n|}\sharp_P Q_n \le  
\frac{1}{|Q_n|}\int_{Q^+_n}f_P(\omega+t)dt  
$$  
and consequently that  
$$  
\lim_{n\to\infty}\frac{1}{|Q_n|}\sharp_P Q_n = \int_\Omega  
f_P(\omega)d\mu(\omega)  
$$  
uniformly in $\omega$.\end{proof}

\medskip  
  
In order to take advantage of the analysis from the theory of tilings,  
let us now relate Delone dynamical systems and tiling dynamical  
systems. To do so, we start with a slight generalization of what can,  
e.g., be found in \cite{sol1}. For the readers convenience we will  
repeat all the necessary definitions. We follow mainly \cite{sol1}  
(see \cite{se} as well) although we rephrase some notions slightly.  
  
A {\it tile} is just a set $T$ that agrees with the closure of its  
interior. A {\it tiling} $S$ is a countable family $S=(T_n)$ of tiles  
with disjoint interiors covering the whole space $\RR^d$.  It is  
sometimes very useful to supply tiles with an additional mark or  
decoration. To do so, consider a finite set $\aA$ called the alphabet  
or the set of {\it decorations}. An {\it $\aA$-decorated tiling} $S$  
consists of a countable family $S=((T_n,a_n))_{n\in\NN}$ such that  
$(T_n)$ is a tiling and the decorations of two tiles agree only if the  
tiles are translates of each other. Following \cite{sol1} we also call  
$a_n$ the {\it type} of the tile $T_n$.

In order to describe a convenient topology for tilings let us first  
ignore decorations. Then, the natural topology provides a suitable  
topology. In fact, let us call  
$$  
\Sigma(S):= \cup_n\partial T_n  
$$  
the shape of the tiling $S=(T_n)$ and $\Sigma$ the shape map acting  
from the set $\CalT$ of all tilings to $\CalF (\RR^d)$. We call the  
initial topology with respect to the shape map the {\it natural  
  topology} on $\CalT$. On the space of $\aA$-decorated tilings  
$\CalT_\aA$ we define the shape map as  
$$  
\Sigma_\aA: \CalT_\aA\to \CalF (\RR^d)^\aA,  
((T_n,a_n))\mapsto(\cup\{ \partial T_n: a_n=a\})_{a\in\aA} .  
$$  
Note that in the case of nonconnected tiles the shape map might  
loose some information. This won't bother us in what follows, since we  
are dealing with convex polygonal tiles.

\begin{definition} {\rm (a)} Let $X\subset \CalT_\aA$ be a set of   
  $\aA$-decorated    
  tilings.  The pair $(X,T)$ is called an {\em $\aA$-tiling dynamical  
    system}, $\aA$-TDS, if $X$ is invariant under the shift  
  $T=(T_t)_{t\in\RR^d}$ and closed in the natural topology.

  {\rm (b)} An $\aA$-TDS $(X,T)$ is said to be of {\em finite type},  
  if there is a finite set of tiles $P$ such that every tile from one  
  of the tilings in $X$ is the translate of one of the tiles in $P$.  
\end{definition}  
  
Next we describe the Voronoi construction. It enables us to pass from  
Delone dynamical to decorated tiling dynamical systems. See the discussion  
in \cite{l1}, Section 2 as well. There, the reader will also find an  
account of the {\it Delone tesselation}, a possibility to  
pass from Delone sets to tilings that goes back to Delone (Delaunay),   
\cite{del}.  
  
Let $\omega\subset\RR^d$ be a Delone set; for $x\in\omega$ define  
\begin{eqnarray*}  
T(x,\omega)&:=&\{ y\in\RR^n:d(x,y)\le d(y,\omega)\}\\  
&=& \{ y\in\RR^n: \| x-y\| \le \| z-y\| \quad (z\in\omega)\} .  
\end{eqnarray*}  
Clearly, $T(x,\omega)$ is the convex hull of finitely many points, a  
polygon, called the Voronoi cell of $\omega$ around $x$. If $\omega$  
is an $(r,R)$-set, it follows that  
$$  
B_{\frac{r}{2}} (x) \subset T(x,\omega)\subset B_{2R}(x) .  
$$  
Moreover,  
$$  
S_\omega:= (T(x,\omega))_{x\in\omega}  
$$  
defines a tiling of $\RR^d$. In this way, we get a TDS $X_\Omega=  
\{ S_\omega :\omega\in\Omega\}$. It is clear that  
$$  
V:\Omega\to X_\Omega, \omega\mapsto S_\omega  
$$  
is continuous and respects translations.  
  
Unfortunately, $V$ doesn't need to be injective. In fact, you will  
easily find different periodic sets (e.g. $\ZZ$ and  
$(\frac14+2\ZZ)\cup (\frac34+2\ZZ)$) that lead to the same Voronoi  
tiling. However, decorations can help to recover the original Delone  
set in the DDSF case. In fact, starting from a DDSF $\Omega$ define  
$$  
\aA:=\{ (\omega\cap B_{2R} (x) -x,  
T(x,\omega)-x)=:a(x,\omega):\omega\in \Omega, x\in\omega\} ,  
$$  
which is a finite set if $\Omega$ is of finite type. Define  
$$  
V:\Omega\to X^\aA_\Omega,  
V_\omega:=((T(x,\omega),a(x,\omega))_{x\in\omega} .  
$$  
It is not hard to see that we can reconstruct $\Omega$ from  
$X^\aA_\Omega$ and that $V$ is an isomorphism of dynamical systems.  
  
Therefore, we can use analogs for Delone dynamical systems of the  
results from \cite{sol1} on ergodic properties of tiling dynamical  
systems.  
  
For a quite different approach to the topology on the set of Delone  
sets we refer to \cite{BHZ}, where Delone sets are identified with the  
sum of delta measures sitting at the points of the Delone set.  Then  
one has the w$^*$-topology on the set of measures at ones disposal,  
providing good compactness properties.  The approach presented here  
has the advantage that a topology is induced on the set of closed  
sets. That can be used to define a topology on (decorated) tilings via  
the shape map.

\section{Groupoids and non commutative integration theory}  
In this section we introduce groupoids and basic notions from Connes  
non-commutative integration theory.

\medskip

We will be concerned with several locally compact topological spaces.  
Given such a space $Z$, we denote the set of continuous functions on  
$Z$ with compact support by $C_c (Z)$. The support of a function in  
$C_c (Z)$ is denoted by $\mbox{supp}(f)$. The topology gives rise to  
the Borel-$\sigma$-algebra. The measurable nonnegative functions with  
respect to this $\sigma$-algebra will be denoted by $\CalF^+(Z)$. The  
measures on $Z$ will be denoted by $\CalM(Z)$.  
  
A set $\CalG$ together with a partially defined associative  
multiplication $\cdot : \CalG^2\subset \CalG\times  
\CalG\longrightarrow \CalG$, and an inversion $-1  
:\CalG\longrightarrow \CalG$ is called a groupoid if the following  
holds:  
  
\begin{itemize}  
\item $(g^{-1})^{-1}=g$ for all $g \in \CalG$,  
\item If $g_1 \cdot g_2$ and $g_2 \cdot g_3$ exist, then $g_1 \cdot  
  g_2 \cdot g_3$ exists as well,  
\item $g^{-1} \cdot g$ exists always and $g^{-1} \cdot g \cdot h = h$,  
  whenever $g \cdot h$ exists,  
\item $h \cdot h^{-1}$ exists always and $g \cdot h \cdot h^{-1} = g$,  
  whenever $g \cdot h$ exists.  
\end{itemize}  
  
A groupoid is called topological groupoid if it carries a topology  
making inversion and multiplication continuous. Here, of course,  
$\CalG\times \CalG$ carries the product topology and $\CalG^2\subset  
\CalG\times \CalG$ is equipped with the induced topology.  
  
A given groupoid $\CalG$ gives rise to some standard objects: The  
subset $\CalG^0 = \{ g \cdot g^{-1} \mid g \in \CalG \}$ is called the  
set of {\it units}. For $g \in \CalG$ we define its {\it range} $r(g)$   
by $r(g) =  
g \cdot g^{-1}$ and its {\it source} by $s(g) = g^{-1} \cdot g$. Moreover,  
we set $\CalG^\omega = r^{-1}(\{ \omega \})$ for any unit $\omega \in  
\CalG^0$. One easily checks that $g \cdot h$ exists if and only if  
$r(h) = s(g)$.  
  
By a standard construction we can assign a groupoid $\CalGomega$ to a  
Delone dynamical system.  As a set $\CalGomega$ is just $\Omega\times  
\RR^n$. The multiplication is given by $(\omega,x)  
(\omega-x,y)=(\omega, x +y)$ and the inversion is given by  
$(\omega,x)^{-1}=(\omega-x,-x)$.  The groupoid operations can be  
visualized by considering an element $(\omega,x)$ as an arrow  
$\omega-x\stackrel{x}{\longrightarrow} \omega$. Multiplication then  
corresponds to concatenation of arrows; inversion corresponds to  
reversing arrows.  
  
Apparently this groupoid $\CalGomega$ is a toplogical groupoid when  
$\Omega$ is equipped with the topology of the previous section and  
$\RR^n$ carries the usual topology.  
  
The groupoid $\CalGomega$ acts naturally on a certain topological  
space $\CalX$. This space and the action of $\CalG$ on it are of  
crucial importance in the sequel. The space $\CalX$ is given by  
$$\CalX=\{(\omega,x)\in \CalG : x\in \omega\}\subset \CalGomega.$$  
In  
particular, it inherits a topology form $\CalGomega$. Two features of  
the topology are given in the following proposition.  
  
\begin{prop} \label{schnitt}{\rm (a)}  $\CalX\subset \CalG$ is closed.  
    
  {\rm (b)} Let $\Oomega$ be an $(r,R)$-system and $\omega\in \Omega$  
  and $x\in \omega$ be arbitrary. Then there exist a neighbourhood $U$  
  of $\omega\in \Omega$ and a continuous function $h: U  
  \longrightarrow B_{\frac{r}{2}} (x)$ with $\omega'\cap  
  B_{\frac{r}{2}} (x) = \{h(\omega')\}$ for every $\omega'\in U$.  
\end{prop}  
\begin{proof} (a) Let $((\omega_\iota, x_\iota)_{\iota\in I})$ be a net  
in $\CalX$ converging to $(\omega,x)\in \CalG$. Thus,  
$\omega_\iota\longrightarrow \omega$ and $x_\iota\longrightarrow x$  
and it remains to show $x\in \omega$. Assume the contrary, i.e.  
$x\notin \omega$. As $\omega$ is closed, there exists a $\delta>0$  
with $B_{\delta}(x)\cap \omega =\emptyset$. Thus,  
$\omega_\iota\longrightarrow \omega$ implies $\omega_\iota\cap  
B_{\frac{\delta}{2}} (x)=\emptyset$ for all $i$ large. But this is a  
contradiction to $x_\iota\longrightarrow x$.  
  
\medskip  
  
(b) By the definition of the topology, there exists a neighbourhood  
$\widetilde{U}$ of $\omega$ with $\omega' \cap B_{\frac{r}{2}}  
(x)\neq\emptyset$ for every $\omega'\in \widetilde{U}$. As $\Oomega$  
is an $(r,R)$-system, $\omega' \cap B_{\frac{r}{2}}(x)$ consists of  
only one element. Denoting this element by $h(\omega')$, we get a  
function $h: \widetilde{U}\longrightarrow B_{\frac{r}{2}}(x)$.  
Continuity of $h$ is now a direct consequence of the definition of the  
topology on $\Omega$. \end{proof}

\begin{remark}{\rm Note that the proof of part (a) does not use that $\Omega$   
is a set of Delone sets. Thus, the corresponding statement remains valid   
whenever $\Omega$ is a subset of  $\CalF(\RR^d)$ which is closed in the   
natural topology.}  
\end{remark}  
  
\medskip  
  
\begin{coro}\label{standard} $C_c (\CalX)=\{f|_\CalX : f\in C_c (\CalG)\}$.   
\end{coro}  
\begin{proof} As $\CalX$ is closed in $\CalG$ be the foregoing  
proposition, this follows by standard arguments involving Uryson's  
Lemma. \end{proof}  
   
\medskip  
  
A key feature of $\CalX$ is its bundle structure. More precisely, we  
have a continuous map $p : \CalX \longrightarrow \Omega$,  
$p((\omega,x))=\omega$ making $\CalX$ into a bundle over $\Omega$ with  
fibres $\CalX^\omega=p^{-1} (\omega)=\{(\omega,p) : p\in  
\omega)\}\simeq \omega\subset\RR^n$.  
  
Now, we can discuss the action of $\CalG$ on $\CalX$. Every $g =  
(\omega,x)$ gives rise to a map $J(g) : \CalX^{s(g)}\longrightarrow  
\CalX^{r(g)}$, $J(g)(\omega-x,p)= (\omega,p +x)$. A simple calculation  
shows that $J(g_1 g_2)=J(g_1) J(g_2)$ and $J(g^{-1})=J(g)^{-1}$,  
whenever $s(g_1)=r(g_2)$. Thus, $\CalX$ is an $\CalG$-space (see  
\cite{lpv}). These $\CalG$-spaces are important objects in Connes  
non-commutative integration theory. They give rise to random  
variables. More precisely we have the following definition.

\begin{definition} Let $\Oomega$ be an $(r,R)$-system.  
   
  {\rm (a)} A choice of measures $\beta: \Omega \to \CalM(\CalX)$ is called  
  a positive random variable with values in $\CalX$ if the map $\omega  
  \mapsto \beta^\omega(f)$ is measurable for every $f \in  
  \CalF^+(\CalX)$, $\beta^\omega$ is supported on $\CalX^\omega$,  
  i.e., $\beta^\omega(\CalX - \CalX^\omega) = 0$, $\omega\in \Omega$,  
  and $\beta$ satisfies the following invariance condition  
\[ \int_{\CalX^{s(g)}} f(J(g)p) d\beta^{s(g)}(p) = \int_{\CalX^{r(g)}}  
f(q) d\beta^{r(g)}(q) \] for all $g \in \CalG$ and $f \in  
\CalF^+(\CalX^{r(g)})$.

{\rm (b)} A map $\Omega\times C_c (\CalX)\longrightarrow \CC$ is called a  
complex random variable if there exist an $n\in \NN$, positive random  
variables $\beta_i$, $i=1,\ldots,n$ and $\lambda_i\in\CC$,  
$i=1,\ldots,n$ with $\beta^{\omega}(f)=\sum_{i=1}^k \beta^\omega_i  
(f)$.  
\end{definition}  
  
We are now heading towards introducing and studying a special random  
variable. This variable is quite important. It will give rise to the  
$\ell^2$-spaces on which the Hamiltonians act. Later we will see that  
these Hamiltonians also induce random variables.  We will need some  
information on the continuous functions on $\CalX$.  
  
\begin{prop}\label{hilf} Let $\Omega$ be an $(r,R)$-system.   
    
  {\rm (a)} Let $g\in C_c (\CalX)$ be given.  Then, the function $\alpha(g)  
  : \Omega \longrightarrow \RR$, $\alpha^\omega (g) = \sum_{p\in  
    \omega} g(\omega,p)$ belongs to $C_c(\CalX)$.  
    
  {\rm (b)} For $g\in \CalF^+(\CalX)$ the function $\alpha(g) : \Omega  
  \longrightarrow \RR$, $\alpha^\omega (g) = \sum_{p\in \omega}  
  g(\omega,p)$ is measurable.

\end{prop}  
  
\begin{proof} (a) We fix $\omega_0\in \Omega$ arbitrary and show  
continuity of $\alpha (g)$ at $\omega_0$.  As $g$ has compact support,  
there exists $s>0$, with $g(\omega,p)=0$ whenever $p\notin B_s$.  
Apparently, $\omega_0\cap B_{s +r}$ is finite. Thus, there exists  
$k\in \NN$ and pairwise different $x_1,\ldots, x_k\in \RR^d$ with  
$\{x_1,\ldots, x_k\}=\omega_0 \cap B_{s + r} $. By Proposition  
\ref{schnitt}, there exist $\widetilde{U}\subset \Omega$ with  
$\omega_0\in \widetilde{U}$ and $h_i : \widetilde{U}\longrightarrow  
B_{\frac{r}{2}}(x_i)$ continuous with $\{h_i  
(\omega)\}=B_{\frac{r}{2}}(x_i)\cap \omega$ for every $\omega\in  
\widetilde{U}$, $i=1,\ldots,k$. As $\Oomega$ is an $(r,R)$-system, the  
$B_{\frac{r}{2}}(x_i)$ are pairwise disjoint.  By definition of the  
topology on $\Omega$, we can find $U\subset \widetilde{U}$ such that  
  
$$\omega \cap B_s\subset \{h_i (\omega) : i=1,\ldots, k\}  
\:\;\mbox{for all $\omega\in U$ }.$$  
(Note that the ball appearing on  
the left hand side of this inclusion has radius strictly less than  
$r+s$.) Thus, for $\omega \in U$, the following holds  
$$\alpha (g)(\omega)=\sum_{p\in \omega \cap B(0,s)} g(\omega,p)=  
\sum_{i=1}^k g(\omega,h_i (\omega))$$  
and the desired continuity  
follows. It is not hard to see that $\alpha (g)$ has compact support.  
  
\medskip  
  
(b) This follows from (a) by standard monotone class arguments. \end{proof}

\medskip  
  
As apparently $\alpha^{\omega} (h)= \alpha^{\omega+x} (h(\cdot -x))$,  
we immediately have the following corollary from part (b) of the  
Proposition.  
  
\begin{coro} \label{corohilf} The map   
$\alpha : \Omega \longrightarrow \CalM (\CalX)$, $\alpha^{\omega} (f)   
= \sum_{p\in \omega} f(p)$ is a random variable with values in $\CalX$.   
\end{coro}

Now, let $\mu$ be a measure on $\Omega$. By (b) of Proposition  
\ref{hilf}, we see that $(\mu\circ \alpha) (g) =\int_\Omega  
\alpha^\omega(g)\, d\mu(\omega)$ exists for every $g\in C_c (\CalX)$.  
The following is the key lemma on integration of random variables.  
  
\begin{lemma}\label{Integration} Let $\mu$ be  $T$-invariant (i.e.   
$\mu(f)=\mu(f(\cdot -t))$ for every $t\in \RR^n$).

  {\rm (a)} Let $\beta$ be a nonnegative random variable. Then $\int_\Omega  
  \beta^{\omega} (F(\omega,\cdot))\,d\mu(\omega)$ does not depend on  
  $F \in \CalF^+ (\CalX)$ provided $F$ satisfies $\int F((\omega+t,  
  x+t )\,dt=1$ for every $(\omega,x)\in\CalX$.  
    
  {\rm (b)} Let $\beta$ be an arbitrary random variable.  $\int_\Omega  
  \beta^{\omega} (F(\omega,\cdot))\,d\mu(\omega)$ does not depend on  
  $F \in \CalF^+ (\CalX)\cap C_c (\CalX)$ provided $F$ satisfies $\int  
  F((\omega+t, x+t )\,dt=1$ for every $(\omega,x)\in\CalX$.  
\end{lemma}    
\begin{proof} (a) This follows from \cite{ni} (see \cite{lpv} for a  
discussion as well).  
  
(b) This is an immediate consequence of (a).  \end{proof}  
  
\medskip  
  
It is instructive to consider a special instance of the lemma.  
Namely, consider $f\in C_c (\RR^n)$. Apparently $f$ gives rise to a  
function $F_f \in C_c (\CalX)$ given by $F_f ((\omega,x))= f(x)$. Now,  
let $\mu$ be an invariant measure on $\Omega$ and $\beta$ a random  
variable. Then, the invariance properties of $\mu$ and $\beta$ show  
that the functional $I : C_c (\RR^n)\longrightarrow \RR$, $I(f)=(\mu  
\circ \alpha) (F_f)$ is translation invariant and positive. By  
uniqueness of the Haarmeasure on $\RR^n$, we infer that there exists a  
constant $\lambda(\beta)$ with $\mu \circ \beta (F_f)= \lambda(\beta)  
\int_{\RR^n} f(t) \,dt$. This shows, in particular, that the integral  
$\int_\Omega \beta^{\omega} (F_f (\omega,\cdot))\,d\mu(\omega)$ does  
not depend on $f\geq 0$ provided $f$ satisfies $\int_{\RR^n} f(t) \, d  
t = 1 $.

Let an invariant measure $\mu$ on $\Omega$ and the random variable  
$\alpha$ as above be given. We can then introduce the space $L^2  
(\CalX, \mu\circ \alpha)$. The bundle structure of $\CalX$ and of  
$\alpha$ suggest, that this space can be considered as a direct  
integral. This means we aim at giving sense to the equation  
\begin{equation}\label{direct}  
L^2 (\CalX,\mu \circ \alpha)=\int_\Omega^{\oplus} \ell^2   
(\CalX^\omega,\alpha^\omega)\,d\mu(\omega).  
\end{equation}  
As the fibres in this direct integral are not constant, we need to be  
careful about the notion of measurability.  More precisely, we need to  
introduce a set $\CalV$ of functions $f$ on $\Omega$ with  
$f(\omega)\in \ell^2 (\CalX^\omega,\alpha^\omega)$, $\omega\in  
\Omega$, satsifying the following properties  
  
\begin{itemize}  
    
\item[(V)] $\CalV$ is a vectorspace under the the obvious operations  
  (i.e. $(f + g)(\omega)= f(\omega) + g(\omega)$ and $(\lambda  
  f)(\omega) = \lambda f(\omega)$).  
    
\item[(M)] $\omega\mapsto \langle f(\omega),g(\omega)\rangle_\omega$  
  is measurable for arbitrary $f,g\in \CalV$. Here,  
  $\langle\cdot,\cdot\rangle_\omega$ is the inner product on $\ell^2  
  (\CalX^\omega,\alpha^\omega)$.  
    
\item[(S)] If $f$ is a function on $\Omega$ with $f(\omega)\in  
  \ell^2(\CalX^\omega,\alpha^\omega)$, $\omega\in \Omega$ and  
  $\omega\mapsto \langle f(\omega),g(\omega)\rangle_\omega$ measurable  
  for every $g\in \CalV$, then $f$ belongs to $\CalV$ as well.  
    
\item[(D)] There exists a countable set $D\subset \CalV$ such that the  
  set $\{d(\omega) : d\in D\}$ is total in $\ell^2  
  (\CalX^\omega,\alpha^\omega)$ for every $\omega \in \Omega$.  
  
\end{itemize}  
  
Such a set will be called a measurable structure on the family  
$(\ell^2 (\CalX^\omega,\alpha^\omega))_{\omega\in \Omega}$.  Note,  
that the condition $(M)$ says that the functions in $\CalV$ have a  
certain measurability property. Condition $(S)$ is a maximality  
assumption.  
  
\medskip  
  
Given the special structure of $\CalX$, we can actually identify  
functions $f$ on $\Omega$ with values in $\ell^2  
(\CalX^\omega,\alpha^\omega)$ with functions on $\CalX$. This will be  
done tacitely in the sequel. There are at least three good canditates  
for measurable structures. They are given as follows:  
  
\begin{itemize}  
    
\item The set $\CalV_1$ consists of all $f : \CalX\longrightarrow \CC  
  $ which are measurable and satisfy $ f(\omega,\cdot)\in \ell^2  
  (\CalX^\omega,\alpha^\omega)$ for every $ \omega\in \Omega$  
    
\item The set $\CalV_2$ consists of all $f : \CalX\longrightarrow \CC$  
  such that $ f(\omega,\cdot)\in \ell^2 (\CalX^\omega,\alpha^\omega)$,  
  for all $\omega\in \Omega$, and $ \omega\mapsto \langle  
  f(\omega,\cdot), F(\omega,\cdot)\rangle_\omega$ is measurable for  
  all $F\in C_c (\CalX)$.

\item Finally, the set $\CalV_3$ is given by all $f :  
  \CalX\longrightarrow \CC$ such that $ f(\omega,\cdot)\in \ell^2  
  (\CalX^\omega,\alpha^\omega)$, for all $\omega \in \Omega$, and $  
  \omega\mapsto \langle f(\omega,\cdot),  
  F(\omega,\cdot)\rangle_\omega$ is measurable for all $F\in C_c  
  (\CalG)$.  
  
\end{itemize}  
  
It is not too hard to check that these are all measurable structures.  
In fact, they are even equal. This is shown next.  
  
\begin{prop}$\CalV_1=\CalV_2=\CalV_3$.   
\end{prop}  
\begin{proof} The equality of $\CalV_2$ and $\CalV_3$ is immediate from  
Corallary \ref{standard}.  
  
\medskip  
  
$\CalV_1\subset \CalV_2$: Let $f\in \CalV_1$ be given. Without loss of  
generality we can assume that $f$ is the characteristic function  
$\chi_M$ of a measurable set $M\subset \CalX$ with $M\cap \omega  
\subset B(0,s)$ for all $\omega \in \Omega$ and a certain $s>0$ not  
depending on $\omega$. As $\CalX$ is both locally compact and  
$\sigma$-compact, its Borel-$\sigma$-algebra is generated by compact  
sets. Thus, it suffices to consider $\chi_K$ with $K\subset \CalX$  
compact. By standard arguments, it then suffices to consider $f\in C_c  
(\CalX)$. For such $f$, measurability follows from Proposition  
\ref{hilf}.  
  
\medskip  
  
$\CalV_2\subset \CalV_1$: Let $f\in \CalV_2$ be given. We have to show  
that $f: \CalX\longrightarrow \CC$ is measurable. By  
$\sigma$-compactness of $\CalX$, it suffices, to find, for every  
$(\omega_0,x_0)\in \CalX$, an open set $U\subset \CalX$ with  
$(\omega_0,x_0)\in U$ such that $f|_U$ is measurable. To provide such  
an $U$, we associate to $(\omega_0,x_0)$ an open set $U_1\subset  
\Omega$ containing $\omega_0$ as well as $h:U_1\longrightarrow  
B_{\frac{r}{2}}(x_0)$ according to Proposition \ref{schnitt}. By  
Uryson's lemma, we can find $U_2\subset U_1$ open containing  
$\omega_0$ and $g\in C_c (\Omega)$ with support contained in $U_1$ and  
$g\equiv 1$ on $U_2$. Moreover, let $s :\RR^d \longrightarrow \RR$ be  
continuous with $s(0)=1$ and support contained in $B_{\frac{r}{4}}$.  
Then, $F:\CalX \longrightarrow \RR$, with $F(\omega,x)= g(\omega)  
s(x-h(\omega))$ whenever $ \omega\in U$ and $F(\omega,x)=0$ otherwise,  
is continuous with compact support. It is immediate that  
  
\begin{equation*}  
  F (\omega,x)= \left\{\begin{array}{r@{\quad:\quad}l}  
 1 &  \omega\in U_2 \;\mbox{and}\; x=h(\omega),\\  
 0 & \omega \in U_2, \;\mbox{and}\; x\neq h(\omega).  
\end{array}\right.  
\end{equation*}  
  
On $U\equiv (U_2\times B_{\frac{r}{4}} (x_0))\cap \CalX$, we then have  
  
$$  
f(\omega,x)=f(\omega,h(\omega))=\langle F(\omega,\cdot),  
f(\omega,\cdot)\rangle_{\omega}$$  
and we infer measurability of $f|_U$  
as $f\in \CalV_2$.

\medskip  
  
It remains to show that $\CalV_1=\CalV_2$ is a measurable structure,  
i.e. satisfies the conditions $(V), (S), (M)$ and $(D)$. Now, $(V)$ is  
clear and $(M)$ is a simple consequence of Corollary \ref{corohilf}.  
  
Moreover, obviously, $C_c (\CalX)$ belongs to $\CalV_1$ and $(S)$  
follows as $\CalV_2\subset \CalV_1$. To show $(D)$, choose for each $q\in  
\QQ^d$ a function $f_q\in C_c (\RR^d)$ with support contained in  
$B_{\frac{r}{2}} (q)$ and $f_q\equiv 1$ on $B_{\frac{r}{4}} (q)$.  
Then, $\{f_q |_\CalX : q\in \QQ^d\}$ has the desired properties.  
\end{proof}

\medskip  
  
Having discussed the appropriate notion of measurability, we can now  
give sense to equation \eqref{direct}. This is the content of the  
following lemma.

\begin{lemma} The map $U: L^2 (\CalX,\mu\circ \alpha)\longrightarrow   
\int_\Omega^{\oplus} \ell^2 (\CalX^\omega,\alpha^\omega)\,d\mu(\omega)$   
with $U(f)(\omega)(x)=f((\omega,x))$  is unitary.   
\end{lemma}  
\begin{proof} By the foregoing proposition, $U(f)$ belongs indeed to  
$\CalV_1$. Direct calculations invoking Fubini's Theorem show that $U$  
is isometric. Thus, $U$ indeed maps into $\int_\Omega^{\oplus} \ell^2  
(\CalX^\omega,\alpha^\omega)\,d\mu(\omega)$ and is injective. To show  
that $U$ is surjective, is suffices to show that its image is dense.  
This can be done as follows: Let $D$ be a dense set of bounded  
functions in $L^2 (\Omega,\mu)$ and $f_q$, $q\in \QQ^d$, as in the  
proof of the foregoing proposition. Then $\{ h :\CalX\longrightarrow  
\CC : h(\omega,x)= g(\omega) f_q (x) \;\:\mbox{for suitable $g´\in D$  
  and $q\in \QQ^d $} \}$ has dense image under $U$. \end{proof}

\medskip  
  
This lemma shows that $L^2 (\CalX,\mu\circ \alpha)$ can be identified  
with $\int_\Omega^{\oplus} \ell^2  
(\CalX^\omega,\alpha^\omega)\,d\mu(\omega)$ in a canonical way.  
  
\medskip  
  
\begin{remark}{\rm In the above considerations, we have introduced $\CalX$ as   
a  tautological bundle over $\Omega$ and then constructed an action of   
$\CalG$ on $\CalX$ as well as a family  $\alpha$ of  measures on $\CalX$. An   
alternative point of view is given as follows: A slight rearrangement of the   
arguments in the proof of Proposition \ref{hilf} shows that    
    $$\alpha^\omega : C_c (\RR^d) \longrightarrow \CC,  
    \;\:\alpha^{\omega} (f)=\sum_{p\in \omega} f(p)$$  
    is continuous in  
    $\omega$ and satisfies an invariance condition. Thus,  
    $\omega\mapsto \alpha^\omega$ is a transverse function on the  
    groupoid $\CalG$ in the sense of Connes  
    non-commutative-integration theory. The space $\CalX$ is then  
    nothing but the ``support'' of $\alpha$. }  
\end{remark}  
  
\section{The von Neuman algebra of random operators}  
In this section we discuss the von Neuman algebra associated to a  
uniquely ergodic dynamical system. Details and proofs will be given in  
\cite{ls2}.  
  
\medskip  
  
Let $\Oomega$ be an $(r,R)$-system and let $\mu$ be an invariant  
measure on $\Omega$.  As there exists a canonical isomorphism between  
$L^2 (\CalX,\mu\circ \alpha)$ and $\int_\Omega^{\oplus} \ell^2  
(\CalX^\omega,\alpha^\omega)\,d\mu(\omega)$, a special role is played  
by operators on $L^2 (\CalX,\mu\circ \alpha)$ which respect this fibre  
structure. More precisely, we consider families  
$(A_\omega)_{\omega\in\Omega}$ of bounded operators $A_\omega :  
\ell^2(\omega,\alpha^\omega)\longrightarrow  
\ell^2(\omega,\alpha^\omega)$. Such a family is called {\it measurable} if  
$\omega\mapsto \langle f(\omega), (A_\omega g)(\omega)\rangle_\omega$  
is measurable for every $f\in \CalV_1$. It is called {\it bounded} if the  
norms of the $A_\omega$ are uniformly bounded. It is called {\it covariant}  
if it satisfies the covariance condition  
\begin{equation}  
H_{\omega+t} = U_t H_\omega U_t^*, \;\omega\in \Omega, t\in \RR^d,  
\end{equation}  
where $U_t :\ell^2(\omega)\longrightarrow \ell^2 (\omega + t)$ is the  
unitary operator induced by translation. Now, we can define  
\begin{equation}  
\CalN:=\{ A=(A_\omega)_{\omega\in\Omega}|A\mbox{  covariant,   
measurable and bounded}\}/\sim,  
\end{equation}  
where $\sim$ means that we identify families which agree $\mu$ almost  
everywhere.  
  
\begin{remark}{\rm  It is possible to define $\CalN$ by requiring seemingly   
weaker conditions. Namely, one can consider families $(H_\omega)$ which are   
essentially bounded and which satisfy the covariance condition almost   
everywhere.   
However, by standard procedures (see \cite{ni,le}), it is possible to show   
that   
each of these families agrees almost everywhere with a family satisfying the   
stronger conditions discussed above.}  
\end{remark}

As is clear from the definition, the elements of $\CalN$ are classes  
of families of operators. However, we will not distinguish too  
pedantically between classes and their representatives in the sequel.  
  
Apparently, $\CalN$ is an involutive algebra under the obvious  
operations.  There is an immediate representation $\pi:  
\CalN\longrightarrow B(L^2 (\CalX,\mu\circ \alpha))$ given by $\pi(A)  
f((\omega,x))= (A_\omega f_\omega)((\omega,x))$. Obviously, $\pi$ is  
injective.  
  
\begin{lemma} $\pi (\CalN)$ is a von Neuman algebra.   
\end{lemma}  
  
\medskip  
  
The elements of $\CalN$ and $\pi(\CalN)$ are called random operators.  
  
\begin{lemma} Let $\mu$ be ergodic and $(A_\omega)\in \CalN$ be selfadjoint.   
Then there exists $\Sigma, \Sigma_{ac},\Sigma_{sc},\Sigma_{pp},\Sigma_{ess}  
\subset \RR$ and a subset  $\widetilde{\Omega}$ of  $\Omega$ of full measure   
such that $\Sigma=\sigma(A_\omega)$ and  $\sigma_{\bullet} (A_\omega)=   
\Sigma_{\bullet}$ for $\bullet=ac,sc,pp,ess$ and $\sigma_{disc}  
(A_\omega)=\emptyset$  for every $\omega\in \widetilde{\Omega}$.   
\end{lemma}  
  
\medskip  
  
Each random operator gives rise to a random variable.  
  
\begin{prop} Let $(A_\omega)\in \CalN$ be given. Then the map $\beta_A :   
\Omega \longrightarrow \CalM(\CalX)$, $\beta_A ^\omega(f)=\tr (A_\omega   
M_f(\omega))$ is a complex random variable.   
\end{prop}  
  
Now, choose a nonnegative $f\in C_c (\RR^n)$ with $\int_{\RR^n} f(x)  
dx =1$.  Combining the previous proposition with Lemma  
\ref{Integration}, we infer that the map  
$$\tau : \CalN\longrightarrow \CC, \;\: \tau(A)=\int_\Omega  
\tr(A_\omega M_f)\,d\mu(\omega)$$  
does not depend on the choice of  
$f$. Important feature of $\tau$ are given in the following lemma.  
  
\begin{lemma} The map $\tau:\CalN\longrightarrow \CC$ is continuous,   
faithful,   
nonegative on $\CalN^+$ and satisfies $\tau(A B) = \tau(B A)$.   
\end{lemma}

Having defined $\tau$, we can now associate a canonial measure  
$\rho_A$ to every selfadjoint $A\in \CalN$.  
  
\begin{definition} For $A\in \CalN$ selfadjoint,  and $B\subset \RR$ Borel   
measurable,   
we set  let $\rho_A(B)\equiv \tau(\chi_B(A))$, where $\chi_B$ is the   
characteristic   
function of $B$.   
\end{definition}

\begin{lemma} Let $A\in \CalN$ selfadjoint be given. Then  $\rho_A$ is a   
spectral measure   
for $A$. In particular, the support of $\rho_A$ agrees with the almost sure   
spectrum   
$\Sigma$ of $A$ and the equality $\rho_A (F)=\tau (F(A))$ holds for every   
bounded   
measurable $F$ on $\RR$.   
\end{lemma}

\begin{theorem}  
  Let $(\Omega,T)$ be a uniquely ergodic, aperiodic DDSF. Let $\mu$ be  
  the unique invariant probability measure. Then $\CalN$ is a factor  
  of type $II_D$, where  
  $$  
  D=lim_{R\to\infty}\frac{\#(\omega\cap B_R(0))}{|B_R(0)|}  
  $$  
  is the {\em density\/} of $\omega$.  
\end{theorem}

\section{Tight binding operators}  
In order to describe the properties of disordered models quantum mechanically   
it  
is common to use a tight binding approach. E.g., a random model is often   
described by an operator on $\ell^2(\ZZ^d)$ consisting of the Laplacian that   
stands for nearest neighbor interactions plus a random potential   
perturbation.   
We search for an analogous description of quasicrystals, introducing the   
following notion that still leaves a lot of flexibility. In comparison with   
the random or almost random case it is again the fact that the space varies   
that makes the  fundamental difference.  
  
\medskip  
  
Related constructions have been introduced by Kellendonk \cite{ke,ke2} and   
later been discussed by Kelledonk/Putnam \cite{kp} and   
Bellissard/Hermann/Zarrouati \cite{BHZ} (see \cite{Ap,put} as well). All   
these works are concerned with $K$-theory. The relevant   $C^\ast$-algebras   
of tight binding operators are then  discussed within the framework of   
discrete goupoids. These groupoids are  transversals of $\CalGomega$   
\cite{kp,BHZ} (see \cite{be2} for discussion of transversals and tight   
binding operators as well). Our discussion below does not use transversals   
and in fact not even groupoids. We rather directly introduce a   
$C^\ast$-algebra of tight binding operators.     
For further details and proofs we refer the reader to \cite{ls1,ls2}.  
  
\begin{definition} Let $\Omega$ be a DDSF. A family $A=(A_\omega)$,   
$A_\omega\in\CalB(\ell^2(\omega))$ is said to be an {\em operator   
(family) of  
finite range\/} if there exists $s>0$ such that  
\begin{itemize}  
\item  
$(A_\omega\delta_x|\delta_y)=0$ if $x,y\in\omega$ and $|x-y|\ge s$.  
\item  
$(A_{\omega+t}\delta_{x+t}|\delta_{y+t})=(A_{\tilde{\omega}}\delta_x|\delta_y)$   
if  
$\omega\cap B_s (x+t)=\tilde{\omega}\cap B_s (x)+t$ and   
$x,y\in\tilde{\omega}$.  
\end{itemize}  
\end{definition}  
This merely says that the matrix elements $A_\omega(x,y)=  
(A_\omega\delta_x|\delta_y)$ of $A_\omega$ only depend on a sufficiently   
large patch around $x$ and  
vanish if the distance between $x$ and $y$ is too large. Since there are  
only finitely many nonequivalent patches, an operator of finite range is   
bounded in the sense that   
$$\| A\|=\sup_{\omega\in\Omega}\| A_\omega\|<\infty.$$  
Moreover it is clear that every such $A$ is covariant and consequently   
$A\in \CalN$ for every invariant measure $\mu$. The completion   
of  
the space of all finite range operators with respect to the above norm is  
a $C^*$--algebra that we denote by $\CalA$. The representations  
$\pi_\omega:A\mapsto A_\omega$ can be uniquely extended to representations  
of $\CalA$ and are again denoted by $\pi_\omega:\CalA  
\to\CalB(\ell^2(\omega))$. We have the following result:  
\begin{theorem}The following conditions on $\Omega$ are equivalent:  
\begin{itemize}  
\item[{\rm (i)}] $(\Omega,T)$ is minimal.  
\item[{\rm (ii)}] For any selfadjoint $A\in\CalA$ the spectrum  
$\sigma(A_\omega)$ is independent of $\omega\in\Omega$.  
\item[{\rm (iii)}] $\pi_\omega$ is faithful for every $\omega\in\Omega$.  
\end{itemize}  
\end{theorem}  
  
Next we relate the ``abstract integrated density of states'' $\rho_A$ to  
the integrated density of states as considered in random or almost random   
models and defined by a volume limit over finite parts of the operator.  
  
Note that for selfadjoint $A\in \CalA$ and bounded $Q\subset\RR^d$   
the restriction $A_\omega|_Q$ defined on $\ell^2(Q\cap\omega)$ has finite   
rank.   
Therefore, the spectral counting function  
$$  
n(A_\omega,Q)(E):=\#\{ \mbox{ eigenvalues of }A_\omega|_Q\mbox{  below  }E\}  
$$  
is finite and  
$\frac{1}{|Q|}n(A_\omega,Q)$ is the distribution function of the measure  
$\rho(A_\omega,Q)$, defined by  
$$  
\langle \rho(A_\omega,Q),\varphi\rangle:=   
\frac{1}{|Q|}\mbox{tr}(\varphi(A_\omega|_Q))\mbox{  for  }\varphi\in C_b(\RR).  
$$  
One of the fundamentals of random operator theory is the existence  
of the infinite volume limit  
$$  
N(E)= \lim_{Q\nearrow\RR^d}\frac{1}{|Q|}n(A_\omega,Q)(E)  
$$  
for every $\omega\in \Omega$. This amounts to the convergence in distribution  
of the measures $\rho(A_\omega,Q)$ just defined. As a first result on weak  
convergence we get:  
\begin{theorem} Let $(\Omega,T)$ be a uniquely ergodic DDSF and $A\in\CalA$   
selfadjoint. Then, for any van Hoove sequence $Q_n$,  
$\rho(A_\omega,Q_n)\to\rho_A\mbox{  weakly as  }n\to\infty.$  
\end{theorem}  
  
\begin{remark}{\rm  This result is analogous to corresponding results   
for random or almost periodic operators as e.g. \cite{shu, be1,be2}.   
It generalizes  results in  Kellendonk's \cite{ke} on tilings associated   
to primitive substitutions.   Its proof uses ideas of the cited  works   
of  Bellissard (see \cite{ke} as well) and of Hof \cite{hof2}.  }  
  
\end{remark}

For strictly ergodic, aperiodoc DDSF, we actually have a much  stronger   
result. Namely, we can show  pointwise and even uniform convergence of   
the corresponding  distribution functions.  Of course, uniform convergence   
follows from vage convergene if the limit is continuous. Thus, let us   
emphasize that in the context of DDSF continuity of the distribution   
function of $\rho$ is wrong in general.  Still  uniform convergence holds.    
Let us  mention that this fits well within the general philosophy that   
everything behaves very uniformly within the reign of quasicrystals.    
All of this  will be discussed in \cite{ls2}.

\end{document}